\documentclass[11pt]{article}
\pdfoutput=1
\usepackage{jcapmod}

\usepackage{amsfonts, amsmath}
\usepackage{booktabs}
\usepackage[english]{babel}
\usepackage{colortbl}
\usepackage{graphicx}
\usepackage{subfig}
\usepackage{hyperref}
\usepackage[load-configurations=astronomy, range-units=brackets, range-phrase=-, per-mode=reciprocal, mode=math]{siunitx}
\usepackage[babel=true]{microtype}

\definecolor{pyBlue}{RGB}{31, 119, 180}
\definecolor{pyRed}{RGB}{214, 39, 40}
\definecolor{pyGreen}{RGB}{44, 160, 44}
\definecolor{pyBlue2}{RGB}{0, 111, 237}
\definecolor{pyRed2}{RGB}{224, 52, 36}

\numberwithin{equation}{section}
\def\beq{\begin{equation}}
\def\eeq{\end{equation}}

\def\d{\mathrm{d}}
\def\Neff{{N_\mathrm{eff}}}
\def\ell{l}

\DeclareSIUnit{\parsec}{pc}
\DeclareSIUnit{\Mpc}{\mega\parsec}
\DeclareSIUnit{\h}{\mathit{h}}

\DeclareRobustCommand{\SkipTocEntry}[4]{}

\begin{document}

\pagenumbering{roman}
\begin{titlepage}
	\baselineskip=15.5pt \thispagestyle{empty}
	
	\bigskip\
	
	\begin{center}
		{\fontsize{20.74}{21}\selectfont \sffamily \bfseries First Constraint on the Neutrino-Induced\\[12pt]Phase Shift in the BAO Spectrum}
	\end{center}
	
	\begin{center}
		{\fontsize{12}{30}\selectfont Daniel Baumann,$^{1}$ Florian Beutler,$^{2,3}$ Raphael Flauger,$^{4}$ Daniel Green,$^{4}$\\[4pt]An\v{z}e Slosar,$^{5}$ Mariana Vargas-Maga\~{n}a,$^{6}$ Benjamin Wallisch$^{7,1}$ and Christophe Y\`{e}che$^{8,3}$} 
	\end{center}
	
	\vspace{0.25cm}
	\begin{center}
		\vskip8pt
		\textsl{$^1$ Institute for Theoretical Physics, University of Amsterdam, Amsterdam, The Netherlands}
		
		\vskip8pt
		\textsl{$^2$ Institute of Cosmology \& Gravitation, University of Portsmouth, Portsmouth, PO1 3FX, UK}

		\vskip8pt
		\textsl{$^3$ Lawrence Berkeley National Laboratory, 1 Cyclotron Road, Berkeley, CA 94720, USA}
		
		\vskip8pt
		\textsl{$^4$ Department of Physics, University of California, San Diego, La Jolla, CA 92093, USA}
		
		\vskip8pt
		\textsl{$^5$ Physics Department, Brookhaven National Laboratory, Upton, NY 11973, USA}
		
		\vskip8pt
		\textsl{$^6$ Instituto de F\'isica, Universidad Nacional Aut\'onoma de M\'exico, Ciudad de M\'exico, Mexico}
		
		\vskip8pt
		\textsl{$^7$ DAMTP, University of Cambridge, Cambridge, CB3 0WA, UK}
		
		\vskip8pt
		\textsl{$^8$ IRFU, CEA, Universit\'e Paris-Saclay, F-91191 Gif-sur-Yvette, France}
	\end{center}
	
	\vspace{1.2cm}
	\hrule \vspace{0.3cm}
	\noindent {\sffamily \bfseries Abstract}\\[0.1cm]
	The existence of the cosmic neutrino background is a fascinating prediction of the hot big bang model. These neutrinos were a dominant component of the energy density in the early universe and, therefore, played an important role in the evolution of cosmological perturbations. The energy density of the cosmic neutrino background has been measured using the abundances of light elements and the anisotropies of the cosmic microwave background~(CMB). A complementary and more robust probe is a distinct shift in the temporal phase of sound waves in the primordial plasma which is produced by fluctuations in the neutrino density and has recently been detected in the~CMB. In this paper, we report on the first constraint on this neutrino-induced phase shift in the spectrum of baryon acoustic oscillations~(BAO) of the BOSS~DR12~data. Constraining the acoustic scale using Planck data while marginalizing over the effects of neutrinos in the~CMB, we find a non-zero phase shift at greater than 95\%~confidence. We also demonstrate the robustness of this result in simulations and forecasts. Besides providing a new test of the cosmic neutrino background, our work is the first application of the BAO~signal to early universe physics and a non-trivial confirmation of the standard cosmological history.
	\vskip10pt
	\hrule
	\vskip10pt
\end{titlepage}

\thispagestyle{empty}
\setcounter{page}{2}
\tableofcontents

\clearpage
\pagenumbering{arabic}
\setcounter{page}{1}

\clearpage
\section{Introduction}
\label{sec:introduction}

A remarkable prediction of the hot big bang model is a thermal background of neutrinos. This cosmic neutrino background~(C$\nu$B) was released one second after the big bang when the rate of neutrino interactions dropped below the expansion rate of the universe and neutrinos were no longer in thermal equilibrium with the rest of the Standard Model. Measuring the~C$\nu$B would establish a window back to this time, when the universe was at nearly nuclear densities. 
 
\vskip4pt
A variety of experiments have been proposed to observe the~C$\nu$B directly~\cite{Weinberg:1962zza, Ringwald:2009bg, Betts:2013uya}. However, because neutrino interactions at low energies are extremely weak, these experiments are very challenging. Cosmological observations, on the other hand, are making an increasingly strong case that the~C$\nu$B has already been detected indirectly. Measurements of the light element abundances and the anisotropies of the cosmic microwave background~(CMB) are sensitive to the expansion rate during the radiation era and, therefore, probe the energy density of the~C$\nu$B. The consistency of the measurements is remarkable, although the interpretation is somewhat sensitive to assumptions about the cosmological model and constraints weaken considerably in some extensions of the $\Lambda$CDM~model.

\vskip4pt
The effect of neutrinos on the perturbations in the primordial plasma has been shown to be a more robust probe of the~C$\nu$B~\cite{Bashinsky:2003tk}. Neutrinos travel near the speed of light~$c$ in the early universe, significantly faster than sound waves in the hot plasma of photons and baryons, and can therefore propagate information ahead of the sound horizon of the plasma. The gravitational influence of this supersonic propagation induces a shift in the phase of the acoustic oscillations that cannot be mimicked by other properties of the plasma~\cite{Bashinsky:2003tk, Baumann:2015rya}. This phase shift has recently been detected in the~CMB~\cite{Follin:2015hya, Baumann:2015rya}, adding to the robustness of the cosmological evidence for the~C$\nu$B.

\vskip4pt
After recombination, photons decoupled from baryons and the sound waves lost their pressure support. The sudden halt to the propagation of these density waves leaves an overdensity of baryons at the scale of the acoustic horizon at recombination. Subsequent gravitational evolution transfers this overdensity to the matter distribution. The power spectrum of galaxies inherits this feature in the form of baryon acoustic oscillations~(BAO). It was recently pointed out that the BAO spectrum should not only exhibit the same phase shift from the supersonic propagation of neutrinos, but that this shift should also be robust to nonlinear gravitational evolution in the late universe~\cite{Baumann:2017lmt}. This makes the phase shift a clean signature of early universe physics. In this paper, we will provide the first constraint on this phase and find it to be consistent with the existence of the cosmic neutrino background with more than 95\%~confidence from the clustering of matter at low redshifts alone. This is achieved by extending the conventional BAO analysis and including the amplitude of the neutrino-induced phase shift as an additional free parameter~\cite{Baumann:2017gkg}. Our analysis also marks the first use of the BAO~feature beyond its application as a standard ruler.

\vskip4pt
The outline of the paper is as follows: In Section~\ref{sec:theory}, we review the theoretical origin of the neutrino-induced phase shift, while in Section~\ref{sec:data}, we construct a template for the phase shift and measure its amplitude in both mock and real data. We find statistically significant evidence for the presence of the phase shift in the BOSS~DR12~dataset, in line with expectations from mocks and forecasts. We conclude, in Section~\ref{sec:conclusions}, with a brief summary of our results and an outlook on future improvements of our measurement. In a set of appendices, we further validate our template-based method (Appendix~\ref{app:tests}) and describe our analysis in configuration space (Appendix~\ref{app:real}).

\section{Theoretical Background}
\label{sec:theory}

The cosmological evidence for the~C$\nu$B relies on our ability to measure the impact of neutrinos on more directly observable quantities. In this section, we will review theoretical aspects of cosmic neutrinos that will allow us to isolate their effect on the BAO spectrum. First, we explain how the energy density in neutrinos depends on the temperature at neutrino decoupling and the spectrum of particles in the Standard Model~(\S\ref{sec:neutrinos}). We then describe how the gravitational influence of the~C$\nu$B leads to a unique shift in the peak locations of the BAO~signal~(\S\ref{sec:phase}) that cannot be mimicked by other effects in the early or late universe.

\subsection{Cosmic Neutrinos}
\label{sec:neutrinos}

Neutrinos interact with the rest of the Standard Model through the weak force. At early times, these interactions were frequent enough to keep the neutrinos in equilibrium with the rest of the primordial plasma. However, at temperatures of a few~\si{MeV}, the neutrino interaction rate dropped below the expansion rate and the neutrinos eventually decoupled. At temperatures of order the mass of the electron, electron-positron annihilation increases the temperature of the photons relative to that of the neutrinos. Assuming perfect decoupling, the conservation of comoving entropy of the plasma implies
\beq
\left(\frac{T_\nu}{T_\gamma} \right)^{3} = \frac{4}{11} \, .	\label{eq:temp_ratio}
\eeq
Since the neutrinos were relativistic before recombination, their energy density at that time can be written as
\beq
\rho_\nu = \frac{7}{8}\Neff \left(\frac{T_\nu}{T_\gamma} \right)^{\!4} \rho_\gamma = \frac{7}{8}\Neff \left(\frac{4}{11} \right)^{\!4/3} \rho_\gamma \, ,	\label{eq:Neff}
\eeq
where~$\rho_\gamma$ is the photon energy density and the parameter~$\Neff$ is the effective number of neutrinos. Accounting for QED~plasma corrections and the fact that neutrinos have not fully decoupled when electrons and positrons annihilated, one finds $\Neff = 3.046$ in the Standard Model~\cite{Mangano:2005cc}. In this sense, measurements of $\Neff > 0$ probe the energy density of the~C$\nu$B and $\Neff \neq 3.046$ would be a signature of physics beyond the Standard Model.

\subsection{Gravitational Signatures}
\label{sec:phase}

While the direct influence of the~C$\nu$B is very weak at late times, neutrinos constituted 41\% of the total energy density of the universe during the radiation-dominated era. Neutrinos therefore had a significant effect on the gravitational evolution at that time, including the expansion of the universe and the evolution of perturbations. Most of the modes observed in the~CMB entered the horizon during the radiation era and thus felt the large influence of these neutrinos. 

\vskip4pt
Holding other cosmological parameters fixed, the neutrino contribution to the expansion history has several effects. First of all, a variation in the neutrino density changes the redshift of matter-radiation equality~$z_\mathrm{eq}$. Moreover, by increasing the expansion rate during the radiation era and, hence, reducing the time over which the sound waves can propagate and diffuse, it also changes the acoustic scale~$r_s$ and the damping scale~$r_D$~\cite{Hou:2011ec}. In a $\Lambda$CDM+$\Neff$ cosmology, the changes to~$z_\mathrm{eq}$ and~$r_s$ are absorbed by other cosmological parameters and do not constrain~$\Neff$. The effect on~$r_D$, on the other hand, is not degenerate with other parameters and drives the current CMB~constraints, $\Neff=3.13^{+0.30}_{-0.34}$~\cite{Ade:2015xua}. However, the modification of~$r_D$ is not a unique property of neutrinos and if we allow other parameters to vary, such as the primordial helium fraction~$Y_p$, constraints on~$\Neff$ degrade significantly\footnote{This difference will become even more pronounced in the next generation of CMB~experiments~\cite{Abazajian:2016yjj}, for which we expect $\sigma(\Neff) = 0.030\text{ and }0.082$ for fixed and varying~$Y_p$, respectively.} to $\Neff = 3.09^{+0.50}_{-0.60}$.

\vskip4pt
A key property of neutrinos is that they do not behave as a fluid, but as a collection of ultra-relativistic free-streaming particles. As a consequence, neutrinos travel at the speed of light while the sound waves in a relativistic fluid, like the photon-baryon fluid, travel at $c_s \approx c/\sqrt{3}$. The supersonic propagation speed of neutrino perturbations creates a characteristic phase shift in the sound waves of the primordial plasma. A useful way to understand the effect is to consider the evolution of a single initial overdensity~\cite{Bashinsky:2002vx, Eisenstein:2006nj}. (For adiabatic fluctuations, the primordial density field is a superposition of such point-like overdensities.) The overdensities of photons, baryons and neutrinos will spread out as spherical shells, while the dark matter perturbation does not move much and will be left behind at the center. Since the neutrinos travel faster than all other perturbations, they induce metric perturbations ahead of the sound horizon~$r_s$ of the acoustic waves of the photon-baryon fluid.\footnote{Note that the size of the sound horizon imprinted in the BAO~signal is slightly larger than the size observed in the CMB~anisotropies. This is because the latter is set at the time of photon decoupling, whereas the former (often denoted~$r_d$) is determined at the slightly later drag epoch when baryons stop being dragged around by the photons.} As was shown in~\cite{Bashinsky:2003tk}, this creates a constant phase shift of the acoustic oscillations in the limit of large wavenumbers. Specifically, during the radiation era, the photon density contrast takes the following schematic form:
\beq
\delta_\gamma(\vec{k}\hskip1pt) \approx A(\vec{k}\hskip1pt) \cos(k r_s + \phi)\, ,
\eeq 
where~$\phi$ is the neutrino-induced phase shift. At linear order in $\epsilon_\nu \equiv \rho_\nu/(\rho_\gamma+\rho_\nu)$, the predicted value of the phase shift is $\phi \approx 0.2 \hskip1pt\pi \,\epsilon_\nu$~\cite{Bashinsky:2003tk, Baumann:2015rya}. This phase shift was recently detected in the CMB~anisotropy spectrum~\cite{Follin:2015hya, Baumann:2015rya} and converted into an independent constraint on the effective number of neutrinos $N_\mathrm{eff}^\phi = 2.3^{+1.1}_{-0.4}$~\cite{Follin:2015hya}. This verified that neutrinos indeed behave as free-streaming particles and cannot be modeled by a relativistic fluid. Of course, any other free-streaming particles will contribute to~$\Neff$ in proportion to their energy density and would lead to $\Neff > 3.046$. This fact makes measurements of~$\Neff$ also a compelling probe of additional relativistic particles beyond the Standard Model of particle physics~\cite{Brust:2013xpv, Chacko:2015noa, Baumann:2016wac}.

\vskip4pt
The same physics that created the CMB~anisotropies also produced the initial conditions for the clustering of matter. After photon decoupling, the sound speed dropped dramatically and the pressure wave slowed down, producing a shell of gas at about~\SI{150}{\Mpc} from the point of the initial overdensity. This shell attracted the dark matter which therefore also developed the same density profile. At late times, galaxies formed preferentially in the regions of enhanced dark matter density and the acoustic scale became imprinted in the two-point correlation function of galaxies. In Fourier space, this is reflected by oscillations whose frequency is determined by the distance of propagation of the primordial sound waves. The same phase shift that was observed in the spectrum of CMB~anisotropies is therefore also expected to be present in the BAO~spectrum. In this paper, we will provide the first measurement of this effect.

\vskip4pt
An interesting feature of the phase shift in the BAO~spectrum is the fact that it is robust to the effects of nonlinear gravitational evolution~\cite{Baumann:2017lmt}. This provides the rare opportunity of extracting a signature of primordial physics that is immune to many of the uncertainties that inflict the modeling of nonlinear effects in large-scale structure observables. The fact that this phase shift should agree both in the~CMB and the~BAO is a highly nontrivial consequence of physics both before and after recombination, and could be an interesting test of exotic extensions of~$\Lambda$CDM. Being a new low-redshift observable, the BAO~phase shift may also help to shed light onto the apparent low-$z$/high-$z$ discrepancies in some cosmological data~\cite{Freedman:2017yms}.

\section{Observational Results}
\label{sec:data}

The analysis of the (isotropic) BAO~signal is usually reduced to the measurement of a single parameter, the BAO~scale. In this work, we consider an extension of the conventional BAO~analysis that takes the information contained in the phase of the spectrum into account~(\S\ref{sec:modifiedAnalysis}). We then provide the first measurement of the neutrino-induced phase shift in the BOSS~data~(\S\ref{sec:analysis}).

\subsection{Modified BAO Analysis}
\label{sec:modifiedAnalysis}

In the previous section, we emphasized that a constant phase shift (at large wavenumbers) is a unique prediction of the cosmic neutrino background in a radiation-dominated universe. In practice, however, we only observe a finite number of modes, some of which evolved primarily during matter domination. This means that recovering all of the accessible information requires an accurate momentum-dependent template for the phase shift that applies to the modes of interest (see Appendix~C of~\cite{Baumann:2017gkg} for further details).

\subsubsection*{Phase template}

\begin{figure}[t]
	\begin{center}
		\includegraphics{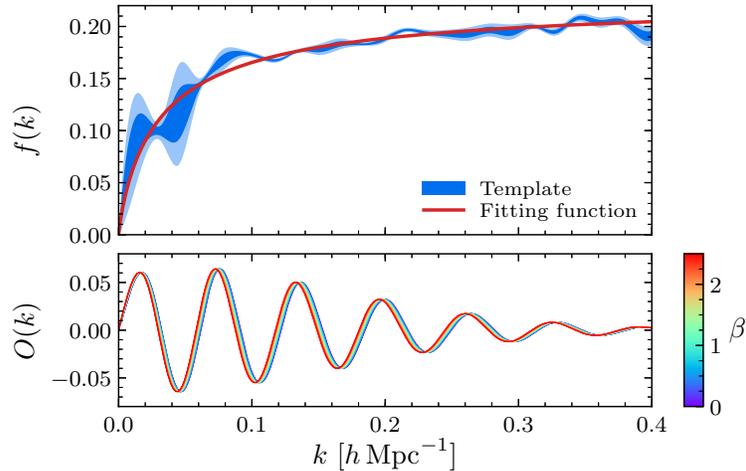}
		\caption{Phase shift induced by free-streaming neutrinos and other light relics. \textit{Top:}~Template of the phase shift $f(k)$~(\textcolor{pyBlue2}{blue}) as defined in~\eqref{eq:phaseTemplate}, with the fitting function~\eqref{eq:phase_fit} shown as the \textcolor{pyRed}{red} curve. The template was obtained numerically in~\cite{Baumann:2017gkg} by sampling the phase shift in 100~different cosmologies with varying free-streaming radiation density. The blue bands indicate the $1\sigma$~and $2\sigma$~contours in these measurements. \textit{Bottom:}~Linear BAO~spectrum~$O(k)$, defined in~\eqref{eq:Pg}, as a function of the amplitude of the phase shift~$\beta$.}
		\label{fig:phaseTemplate}
	\end{center}
\end{figure}

To isolate the BAO~spectrum, we define the following decomposition of the galaxy power spectrum:
\begin{align}
P_g(k) \equiv P^\mathrm{nw}(k) [1+O(k)]\, ,	\label{eq:Pg}
\end{align}
where~$P^\mathrm{nw}(k)$ denotes the smooth (`no-wiggle') spectrum and $O(k) \approx A_\mathrm{w}(k) \sin(k r_d+\phi(k))$ is the BAO~(`wiggle') spectrum, with~$r_d$ being the sound horizon at the drag epoch. Since the phase shift~$\phi(k)$ is robust to nonlinearities, it was numerically extracted in~\cite{Baumann:2017gkg} using the linear spectra~$P^\mathrm{nw}_\mathrm{lin}$ and~$O_\mathrm{lin}$. The phase shift (relative to $\Neff=0$) can be written as
\beq
\phi(k) \equiv \beta(\Neff) f(k) \, ,	\label{eq:phaseTemplate}
\eeq
where~$\beta$ is the amplitude of the phase shift and $f(k)$~is a function that encodes its momentum dependence. Theoretically, we expect~$f(k)$ to approach a constant for $k \to \infty$ in order to match the behavior in a radiation-dominated universe. The $k$-dependence of the phase template, however, will be important for observable scales in a realistic cosmology. The amplitude is proportional to the fractional neutrino density, $\epsilon_\nu(\Neff) \approx \Neff/(4.4+\Neff)$, and we have chosen the normalization so that $\beta = 0\text{ and }1$ correspond to $\Neff = 0\text{ and }3.046$, respectively. We note that the parameter~$\beta$ is a nonlinear function of~$\Neff$ that asymptotes to $\beta \to 2.45$ for $\Neff \to \infty$. As neutrinos become the dominant source of energy density in the universe, adding more neutrinos does not change the phase shift. The template~$f(k)$ is shown in Fig.~\ref{fig:phaseTemplate} and is well approximated by the fitting function
\beq
f(k) = \frac{\phi_\infty}{1+(k_\star/k)^\xi} \, ,	\label{eq:phase_fit}
\eeq
where $\phi_\infty = 0.227$, $k_\star = \SI{0.0324}{\h\per\Mpc}$ and $\xi = 0.872$. This template is essentially independent of changes to the BAO scale $r_d$, for example due to changes in the dark matter density.

\subsubsection*{Model of the BAO spectrum}

The observed BAO~spectrum receives various nonlinear corrections. We model these contributions as in the standard BAO~analysis, e.g.~\cite{Beutler:2016ixs}, but now introduce the amplitude of the phase shift~$\beta$ as an additional free parameter, i.e. we write the nonlinear BAO~spectrum as 
\beq
O(k) \equiv O^\mathrm{fid}_\mathrm{lin}\big(k / \alpha + (\beta-1) f(k)/r_d^\mathrm{fid} \big) \,\mathrm{e}^{-k^2\Sigma_\mathrm{nl}^2 /2}\, ,	\label{eq:ONL}
\eeq
where~$O^\mathrm{fid}_\mathrm{lin}(k)$ and~$r_d^\mathrm{fid}$ are the linear BAO~spectrum and the BAO~scale in the fiducial cosmology, which is chosen to be the same as in~\cite{Beutler:2016ixs}. The exponential factor in~\eqref{eq:ONL} describes the nonlinear damping of the BAO~signal after reconstruction~\cite{Eisenstein:2006nk, Padmanabhan:2012hf}. The parameter~$\alpha$ captures the change in the apparent location of the BAO~peak due to changes in the acoustic scale and the angular projection, 
\beq
\alpha(\Neff) = \frac{D_V(z)\,r_d^\mathrm{fid}}{D_V^\mathrm{fid}(z)\,r_d}\, , \quad \text{with} \quad D_V(z) = \left[(1+z)^2D^2_A(z)\frac{cz}{H(z)}\right]^{1/3}\, ,
\eeq
where~$D_A(z)$ and~$H(z)$ are the angular diameter distance and the Hubble rate at redshift~$z$, respectively. In Appendix~\ref{app:tests}, we show that this model is effectively unbiased in the sense that we recover $\beta \approx 0$ for a universe with $\Neff = 0$ even when we assume a fiducial model with $\Neff=3.046$. Moreover, given the template~\eqref{eq:phase_fit}, the modeling is robust to the precise method for extracting~$O^\mathrm{fid}_\mathrm{lin}(k)$ and we will therefore use the same method as~\cite{Beutler:2016ixs}.

\vskip4pt
We model the nonlinear broadband spectrum in each redshift bin as
\beq
P^\mathrm{nw}(k) = B^2 P^\mathrm{nw}_\mathrm{lin}(k) F(k,\Sigma_s) + A(k)\, .	\label{eq:smiso}
\eeq
This includes two physical parameters: a linear bias parameter,~$B$, and a velocity damping term arising from the nonlinear velocity field (``Fingers of God''),
\beq
F(k,\Sigma_s) = \frac{1}{(1 + k^2\Sigma_s^2/2)^2}\, .	\label{eq:FoG}
\eeq
In addition, we have introduced the polynomial function
\beq
A(k) = \frac{a_{1}}{k^3} + \frac{a_{2}}{k^2} + \frac{a_{3}}{k} + a_{4} + a_{5}k^2\, ,	\label{eq:fs_poly}
\eeq
whose coefficients~$a_n$ will be marginalized over. This polynomial does not represent a physical effect, but removes any residual information that is not encoded in the locations of the peaks and zeros of the BAO spectrum. With such a marginalization over broadband effects, our \mbox{$\alpha$-$\beta$}~parameterization contains essentially all of the information of the $\Lambda$CDM+$\Neff$ cosmology available in the BAO~spectrum~\cite{Baumann:2017gkg}. The free parameters in this model will be fit independently in each redshift bin.

\subsection{Application to BOSS Data}
\label{sec:analysis}
 
We have applied our method to the BAO~signal of the final data release~(DR12) of the Baryon Oscillation Spectroscopic Survey~(BOSS); see~\cite{Alam:2015mbd}. The survey covers~\SI{10252}{deg^2} of the sky and contains \num{1198006}~galaxies with spectroscopic redshifts in the range $0.2 < z < 0.75$. The sample is described in detail in~\cite{Reid:2015gra}.

\vskip4pt
In our model, the measured galaxy power spectrum is described by two cosmological parameters, $\alpha$~and~$\beta$, and a number of nuisance parameters. Except for~$\beta$, all free parameters are redshift dependent and will be fit independently in each of the two separate redshift bins, ($0.2 < z_1 < 0.5$) and ($0.5 < z_3 < 0.75$).\footnote{The middle redshift bin ($0.4 < z_2 < 0.6$), which was used in the BOSS~DR12~analysis, carries little additional information on the BAO~signal since it overlaps with the other two bins.} In total, our fit to the power spectrum in the range $\SI{0.01}{\h\per\Mpc} < k < \SI{0.3}{\h\per\Mpc}$ therefore has 21~free parameters:
\beq
\beta,\, \alpha_{z_1},\, \alpha_{z_3};\, \{ B_{\mathrm{NGC},z},\, B_{\mathrm{SGC},z},\, \Sigma_{s,z},\,\Sigma_{\mathrm{nl},z},\, a_{n,z} \}_{z_1,z_3}\,, 
\eeq
where we have allowed for independent bias parameters in the North Galactic Cap~(NGC) and South Galactic Cap~(SGC) as in~\cite{Beutler:2016ixs}. Throughout the analysis, we employ the galaxy power spectrum after BAO~reconstruction~\cite{Eisenstein:2006nk, Padmanabhan:2012hf}; previous works suggest this choice will not induce a bias in the \mbox{$\alpha$-$\beta$}~plane at BOSS~uncertainties (e.g.~\cite{Mehta:2011xf, Xu:2012fw, Vargas-Magana:2016imr, Baumann:2017lmt, Ding:2017gad, Sherwin:2018wbu}).

\vskip4pt
To explore the BAO~likelihood function, we use the Python-based, affine-invariant ensemble sampler \texttt{emcee}~\cite{ForemanMackey:2012ig} for Markov chain Monte Carlo. The convergence is determined with the Gelman-Rubin criterion~\cite{Gelman:1992zz} by comparing eight separate chains and requiring all scale-reduction parameters to be smaller than $\epsilon = 0.01$. We impose flat priors on all parameters, in particular~$\beta$.\footnote{We point out that the choice of a flat prior on~$\beta$, rather than~$\Neff$, weakens the statistical significance of the $\beta>0$ constraint compared to the analyses in the~CMB which use~$\Neff$. In other words, a flat prior on~$\Neff$ would lead to stronger constraints on the phase shift and, therefore the~C$\nu$B.} We require the $\alpha_z$~parameters to be between $0.8$~and~$1.2$, and the damping scales, $\Sigma_{s,z}$~and~$\Sigma_{\mathrm{nl},z}$, to be between $0$~and~\SI{20}{\per\h\Mpc}, while no explicit priors are employed for the bias parameters~$B_{i,z}$, the phase parameter~$\beta$ or the polynomial terms~$a_{n,z}$. Our goal is to constrain the new parameter~$\beta$, while marginalizing over all other parameters.

\subsubsection*{Validation using mock catalogs}

Before applying our analysis pipeline to the BOSS~data, we validated our method through likelihood-based forecasts and on 999~MultiDark-Patchy mock catalogs~\cite{Kitaura:2015uqa}, which have been created for the BOSS~DR12~analysis. The Patchy mock catalogs have been calibrated to an N-body simulation-based reference sample using analytical-statistical biasing models. The reference catalog is extracted from one of the BigMultiDark simulations~\cite{Klypin:2014kpa}. The mock catalogs have a known issue with overdamping of the~BAO, making the signal for the traditional~BAO approximately 30\%~weaker~\cite{Beutler:2016ixs}. We therefore forecast the mocks and the real data separately, taking these differences into account. For the mock forecasts, we used $\Sigma_\mathrm{nl}=\SI{7}{\per\h\Mpc}$ as the fiducial value of the nonlinear damping scale.

\begin{figure}[t]
	\begin{center}
	 	\includegraphics{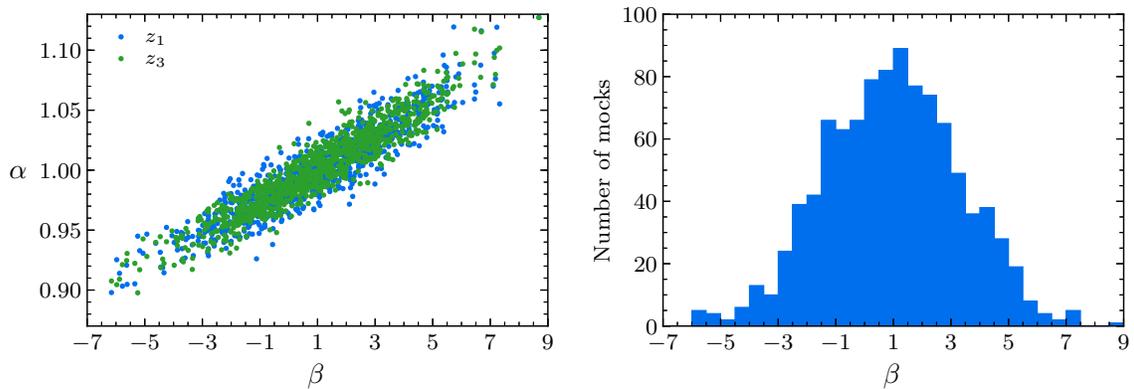}
		\caption{Validation of the Fourier-space analysis using mock catalogs. We compute the maximum-likelihood~(ML) values for the BAO~frequency parameter~$\alpha$ and phase shift amplitude~$\beta$ in 999~mock catalogs~\cite{Kitaura:2015uqa} to validate our analysis pipeline. \textit{Left:}~The distribution of ML~values in the $\alpha$-$\beta$ plane for the two redshift bins~$z_1$ and~$z_3$ exhibits the expected degeneracy. \textit{Right:}~The marginalized one-dimensional distribution of ML~values for~$\beta$ yields $\beta = 1.0 \pm 2.4$ which is consistent with the constraints expected from a likelihood-based forecast.}
		\label{fig:mocks}
	\end{center}
\end{figure}

\vskip4pt
An appealing feature of using the mock catalogs is that we can check that the performance expected from our forecasts~\cite{Baumann:2017gkg} is reproduced by the distribution of maximum-likelihood points across the catalog. Figure~\ref{fig:mocks} confirms that the distributions for the parameters~$\alpha$ and~$\beta$ are indeed in good agreement with the fiducial value of $\beta=1$. A Gaussian fit to the distribution of maximum-likelihood values yields $\beta = 1.0 \pm 2.4$ ($\alpha_{z_1} = 1.000 \pm 0.035$, $\alpha_{z_3} = 1.000 \pm 0.035$), which is consistent with the value found from a likelihood-based forecast as in~\cite{Baumann:2017gkg}, $\sigma(\beta) = 2.1$.

\vskip4pt
As seen in the left panel of Fig.~\ref{fig:mocks}, there is a strong degeneracy between the effects of the parameters~$\alpha$ and~$\beta$. The origin of this degeneracy is easy to understand. If the only well-determined quantity in the data were the position of the first peak in the BAO~spectrum, there would be a perfect degeneracy between phase and frequency determination. In reality, several peaks and troughs are present in the data which breaks the perfect degeneracy and allows the parameters~$\alpha$ and~$\beta$ to be constrained independently. However, one still expects them to remain significantly correlated, partly because the peaks are measured with decreasing accuracy due to damping. Since this degeneracy is a limiting factor in the determination of~$\beta$, we anticipate a significant improvement in the constraint on~$\beta$ when the degeneracy with~$\alpha$ is broken with additional data. Below we will see that this is indeed the case.

\subsubsection*{Analysis of BOSS DR12 data}

\begin{figure}[t!]
	\begin{center}
		\includegraphics{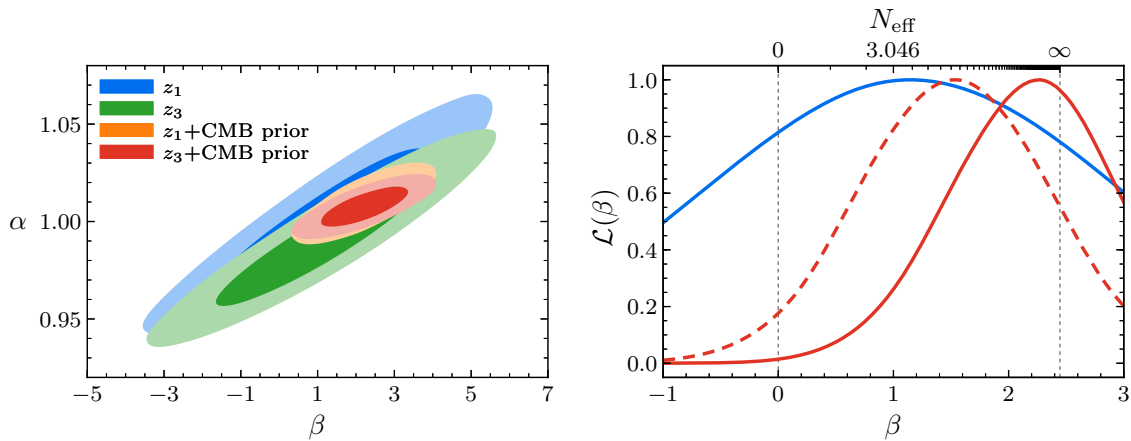}
		\caption{Observational constraints on the amplitude of the phase shift~$\beta$. \textit{Left:}~Contours showing $1\sigma$~and $2\sigma$~exclusions in the $\alpha$-$\beta$ plane for the two redshift bins~$z_1$ and~$z_3$ from our Fourier-space analysis of the BOSS~DR12~data, both from the BAO~data alone and after imposing a CMB~prior on the BAO~frequency parameter~$\alpha$. The degeneracy between the parameters~$\alpha$ and~$\beta$ is clearly visible. By imposing a prior on~$\alpha$ from the~CMB, we restrict the values of the BAO~frequency, or equivalently the BAO~scale, to be consistent with observational constraints from the Planck satellite. \textit{Right:}~One-dimensional posterior distributions of~$\beta$ without~(\textcolor{pyBlue2}{blue}) and with~(\textcolor{pyRed2}{red}) the $\alpha$-prior from Planck for the combined redshift bins. The dashed line is the result after marginalizing over the lensing amplitude~$A_L$, which is a phenomenological parameter that exhibits a large fluctuation in the cosmology inferred from the Planck data. Even in this case, we exclude $\beta=0$ at more than 95\%~confidence.}
		\label{fig:2Dlike}
	\end{center}
\end{figure}

We then applied our analysis pipeline to the BOSS~DR12 dataset, extending the standard BAO~analysis presented in~\cite{Beutler:2016ixs, Alam:2016hwk} by including the phase shift parameter~$\beta$. Figure~\ref{fig:2Dlike} shows the posterior distribution for the parameters~$\beta$ and~$\alpha_{z_1}, \alpha_{z_3}$. The measured $\alpha$-values are in good agreement with those found in~\cite{Beutler:2016ixs}, but the errors have increased due to the degeneracy with~$\beta$. We find $\alpha_{z_1} = 1.001\pm0.025$, $\alpha_{z_3} = 0.991\pm 0.022$ and $\beta = 1.2 \pm 1.8$. Accounting for the linear galaxy bias measured in~\cite{Beutler:2016ixs}, these results are in good agreement with forecasts for the data based on~\cite{Baumann:2017gkg}, $\sigma(\alpha_{z_1})=0.021$, $\sigma(\alpha_{z_3})=0.019$ and $\sigma(\beta) = 1.5$. A similar level of agreement between forecasts and actual performance was obtained for the measurement of~$\alpha$ in the conventional BAO~analysis of BOSS~DR12~\cite{Beutler:2016ixs}.

\vskip4pt
While the phase shift is naturally described in Fourier space, the measurement of the BAO~scale is often depicted as the determination of the BAO~peak location in configuration space~\cite{Ross:2016gvb, Vargas-Magana:2016imr}. In configuration space, the phase shift modifies the shape of the BAO~peak, moving correlations around the peak position from small to large scales. As described in Appendix~\ref{app:real}, we have also incorporated this change into the configuration-space analysis of the BAO~signal. The resulting constraint on the amplitude of the phase shift is $\beta = 0.4 \pm 2.1$, which is statistically consistent with the result of the Fourier-space analysis. While the change to the BAO~peak is simply the inverse Fourier transform of the phase shift, the broadband modeling and peak isolation in configuration and Fourier space are distinct, and the agreement between the two analyses confirms that a comparable constraint can also be obtained in configuration space.

\subsubsection*{Adding a CMB prior}

The BAO-only constraint on~$\beta$ is limited by the degeneracy with~$\alpha(z)$ due to the finite range of wavenumbers. This degeneracy arises because it is hard to extract the phase of an oscillation with an unknown frequency. However, in a given cosmology, $\alpha(z)$~is determined by a few cosmological parameters that are measured precisely by other means, even when marginalizing over the~C$\nu$B. Furthermore, the neutrino-induced phase shift is a non-trivial signature of the~C$\nu$B and is distinct from our knowledge of any other cosmological parameters. We are therefore interested in constraining the neutrino-induced phase shift in the BAO~signal assuming a background cosmology that is consistent with the Planck CMB~constraints. By construction, this restriction on~$\alpha(z)$ carries no information about~$\beta$ since it only limits the frequency of the baryon acoustic oscillations to lie within observational uncertainties. We infer the prior on~$\alpha(z)$ from the Planck~2018 temperature and polarization data\hskip1pt\footnote{We use the low-multipole ($2 \leq \ell \leq 29$) temperature and High Frequency Instrument~(HFI) polarization data, and the high-multipole ($30 \leq \ell \leq 2508$) \texttt{plik} cross half-mission temperature and polarization spectra~\cite{Aghanim:2018eyx}. In~``TT-only'', we omit the high-$\ell$ polarization spectra. The $\Lambda$CDM+$\Neff$+$A_L$ prior cosmology is evaluated on Planck~2015 data with the same specifications, but employing Low Frequency Instrument~(LFI) polarization data~\cite{Ade:2015xua}.}~\cite{Aghanim:2018eyx} while marginalizing over any additional cosmological information (including all effects of~$\Neff$). If available, we directly employ the Markov chains supplied by the Planck collaboration, which were computed using \texttt{CAMB}~\cite{Lewis:1999bs} and \texttt{CosmoMC}~\cite{Lewis:2002ah} with the publicly released priors and settings. In particular, for the $\Lambda$CDM+$\Neff$+$A_L$ prior cosmology, we sample the data using the same codes and priors. At each point in the Monte Carlo Markov chains obtained from the Planck likelihood for a certain background cosmology, we compute the values of~$\alpha_{z_1}$ and~$\alpha_{z_3}$ associated with the given set of cosmological parameters. In this way, we infer the two-dimensional (Gaussian) posterior for $\alpha_{z_1}$-$\alpha_{z_3}$. We confirmed on the mock catalogs that a Gaussian prior with the expected mean values and the Planck $\Lambda$CDM+$\Neff$ covariance matrix results in an unbiased determination of $\beta = 1.00 \pm 0.85$ (see also Appendix~\ref{app:tests} for the equivalent forecasts). On the data, we impose the Planck posterior on~$\alpha$ by importance-sampling our BAO-only Monte Carlo Markov chains.

\vskip4pt
The right panel of Fig.~\ref{fig:2Dlike} shows the marginalized posterior distributions for the parameter~$\beta$. We see that including the $\alpha$-posterior from the Planck $\Lambda$CDM+$\Neff$ chains sharpens the distribution significantly. Having obtained a constraint on the phase amplitude of $\beta = 2.22\pm 0.75$, we want to evaluate the statistical significance of an exclusion of $\beta = 0$, corresponding to no phase shift and no free-streaming neutrinos. For this purpose, we extract the fraction of Monte Carlo samples which have $\beta>\beta_0$. To be cautious about the small bias found in the likelihood-based forecasts when inferring the posterior of~$\beta$ from mock BOSS~data with $N_\mathrm{eff}^\mathrm{in}=0.0$ (see Appendix~\ref{app:tests}), we use $\beta_0 = 0.27$ instead of $\beta_0 = 0$.\footnote{We also checked that the computation based on likelihood ratios leads to essentially the same confidence levels, which is expected since the posterior distributions are very close to Gaussian.} In this and other aspects of the analysis, we have therefore made intentionally conservative choices. The measurement of $\beta = 2.22\pm 0.75$ consequently corresponds to an exclusion of $\beta = 0$ at greater than 99\%~confidence. The statistical error of this result is in good agreement with the forecasted value of $\sigma(\beta) = 0.77$. On the other hand, the central value is more than a $1\sigma$~fluctuation away from the expected Standard Model value $\beta = 1$. Any upward fluctuation adds to the confidence of our exclusion, provided that it is simply a statistical fluctuation. %
\begin{table}
	\centering
	\begin{tabular}{l S[table-format=1.2, separate-uncertainty, table-figures-uncertainty=2]}
			\toprule
		Prior Cosmology						& {$\beta$}		\\
			\midrule[0.065em]
		None (BAO-only)						& 1.2  \pm 1.8	\\
			\midrule[0.02em]
			\rowcolor[gray]{.9}
		$\Lambda$CDM+$\Neff$				& 2.22 \pm 0.75	\\
		$\Lambda$CDM						& 2.05 \pm 0.70	\\
			\midrule[0.02em]
		$\Lambda$CDM+$\Neff$ (TT-only)		& 2.2  \pm 1.0	\\
		$\Lambda$CDM (TT-only)				& 2.16 \pm 0.87	\\
			\midrule[0.02em]
		$\Lambda$CDM+$\Neff$+$A_L$ (2015)	& 1.53 \pm 0.83	\\
		$\Lambda$CDM+$A_L$					& 1.30 \pm 0.76	\\
			\bottomrule 
	\end{tabular}
	\caption{Observational constraints on the amplitude of the phase shift~$\beta$. We infer these constraints on the phase shift from the BOSS~DR12~data with and without a Planck prior on the BAO~scale, assuming various underlying cosmologies. Our baseline result uses the $\Lambda$CDM+$\Neff$ prior, marginalizing over all of the effects of~$\Neff$ in the~CMB. We see that this result is robust to including or excluding~$\Neff$ and~$A_L$ in the prior cosmology. Finally, we show that the large central value of $\beta$ also appears when only using temperature~(`TT-only') spectra and is therefore not solely a consequence of the polarization data.}
	\label{tab:bao+priorResults}
\end{table}
We tested the stability of this upward fluctuation to changes in the cosmological model and the CMB~likelihood (see Table~\ref{tab:bao+priorResults}). The statistical significance of the result is largely insensitive to the choice of cosmology and likelihood. The largest deviation from~$\Lambda$CDM within the Planck data alone is the preference for a larger lensing amplitude~$A_L$~\cite{Aghanim:2016sns}. To estimate the impact of this upward fluctuation on our analysis, we marginalized over~$A_L$ in the implementation of the $\alpha$-prior. The dashed posterior curve in Fig.~\ref{fig:2Dlike} shows the result obtained from the $\Lambda$CDM+$\Neff$+$A_L$ prior cosmology, which corresponds to $\beta = 1.53 \pm 0.83$. We see that marginalizing over~$A_L$ indeed brings the central value of~$\beta$ into closer agreement with $\beta=1$, suggesting that part of our large central value is due to a known upward fluctuation of the Planck data. Having said that, even with this marginalization, we find a positive phase shift, $\beta>0$, at greater than 95\%~confidence.\footnote{Note that we marginalized over~$A_L$ because it experiences a large fluctuation in the Planck data, which is why the statistical significance of the corresponding result should not be compared to the results of our blind analysis.} Finally, we have also implemented the CMB~prior in the configuration-space analysis, obtaining results that are broadly consistent with those in Fourier space. For example, we find $2.55 \pm 0.80$ when including the $\Lambda$CDM+$\Neff$ prior. In summary, while the precise significance of the non-zero phase shift depends on the implementation of the CMB~prior, the exclusion of $\beta=0$ at greater than 95\%~confidence is stable to all choices of the prior that we have considered.

\section{Conclusions and Outlook}
\label{sec:conclusions}

In this paper, we have reported on the first constraint on the neutrino-induced phase shift in the BAO~spectrum. This is the first evidence for the cosmic neutrino background in the clustering of galaxies and the first application of the BAO~signal beyond its use as a standard ruler.

\vskip4pt
To extract the phase information, we modified the conventional BAO data analysis by allowing the amplitude of the phase shift to be an additional free parameter. We determined this new parameter to be non-zero at greater than 95\%~confidence, even allowing for very conservative marginalization over corrections to the broadband spectrum. Our result is a nontrivial confirmation of the standard cosmological model at low redshifts and a proof of principle that there is additional untapped information in the phase of the BAO spectrum, both for the cosmic neutrino background and beyond. Since this phase information is protected from the effects of nonlinear gravitational evolution~\cite{Baumann:2017lmt}, it is a particularly robust probe of early universe physics.
\begin{figure}[htb]
	\begin{center}
		\includegraphics{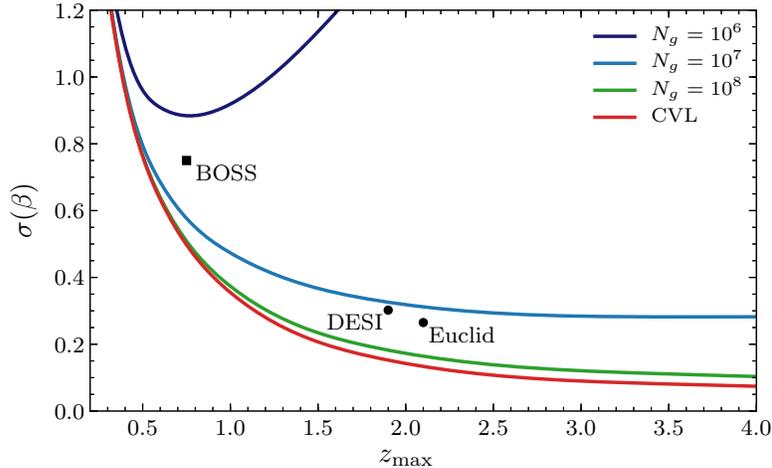}
		\caption{Current and future constraints on the amplitude of the phase shift~$\beta$. The lines are forecasted constraints, which include a CMB~prior on the BAO~scale parameter~$\alpha$ from Planck, as a function of the maximum redshift~$z_\mathrm{max}$ and the number of objects~$N_g$ of a cosmological survey observing a sky fraction of $f_\mathrm{sky}=0.5$ (see~\cite{Baumann:2017gkg} for details). Shown is also the cosmic variance limit~(CVL). The square indicates the result obtained in this work. The circles mark projected constraints for~DESI and Euclid assuming~$z_\mathrm{max}$ to be given by the largest redshift bin used to define the survey in~\cite{Font-Ribera:2013rwa}.}
		\label{fig:conclusions}
	\end{center}
\end{figure}

\vskip4pt
While we have demonstrated that BOSS~data already place an interesting constraint on this phase, a number of galaxy surveys are planned over the next decade which have the potential to significantly improve on our measurement of the neutrino background (see Fig.~\ref{fig:conclusions}). The Dark Energy Spectroscopic Instrument~(DESI), for example, should be sensitive to the~C$\nu$B at more than~$3\sigma$~\cite{Baumann:2017gkg}, making the constraint on the BAO~phase shift more comparable to current limits from the~CMB~\cite{Follin:2015hya}. Combining Euclid with a prior from a next-generation CMB~experiment would allow a $5\sigma$~detection of the~C$\nu$B. Moreover, having shown that there is valuable information in the phase of the BAO~spectrum, we should ask what else can be learned from it beyond the specific application to light relics. As the observed BAO~feature is the result of the combined dynamics of the dark matter and baryons, it is broadly sensitive to new physics in these sectors. The BAO~phase shift is one particularly clean probe of this physics and we hope that our work will inspire new ideas for exploring the early universe at low redshifts.

\vskip20pt
\paragraph{Acknowledgements}
D.\,B.~is grateful to the Leung Center for Cosmology and Particle Astrophysics at the National Taiwan University and the Institute for Advanced Study for hospitality while this work was being completed, and is supported by a Vidi grant of the Netherlands Organisation for Scientific Research~(NWO) that is funded by the Dutch Ministry of Education, Culture and Science~(OCW). F.\,B.~is a Royal Society University Research Fellow. R.\,F.~is supported in part by the Alfred~P.\ Sloan Foundation, the Department of Energy under Grant No.~DE-SC0009919, a grant from the Simons Foundation/SFARI~560536, and by NASA under Grant No.~17-ATP17-0193. D.\,G.~thanks the University of California, Berkeley for hospitality while this work was being completed. A.\,S.~thanks the Cosmoparticle Hub at University College London for hospitality during periods where parts of this manuscript were prepared. M.\,V.~is partially supported by Programa de Apoyo a Proyectos de Investigaci\'on e Innovaci\'on Tecnol\'ogica~(PAPITT) No.~IA102516 and No.~IA101518, Proyecto Conacyt Fronteras No.~281 and Proyecto LANCAD-UNAM-DGTIC-319. B.\,W.~is grateful to the CERN~theory group for its hospitality and acknowledges support by a Cambridge European Scholarship of the Cambridge Trust, a Research Studentship Award of the Cambridge Philosophical Society, an STFC~Studentship and a Visiting PhD~Fellowship of the Delta-ITP consortium, a program of~NWO.

This work is based on observations obtained by the Sloan Digital Sky Survey~III~(\mbox{SDSS-III}, \href{http://www.sdss3.org/}{http:/\!/www.sdss3.org/}). Funding for SDSS-III has been provided by the Alfred~P.\ Sloan Foundation, the Participating Institutions, the National Science Foundation and the U.S.~Department of Energy Office of Science. This research also partly uses observations obtained with the Planck satellite (\href{http://www.esa.int/Planck}{http:/\!/www.esa.int/Planck}), an ESA~science mission with instruments and contributions directly funded by ESA~Member States, NASA and Canada. Some parts of this work were undertaken on the COSMOS~Shared Memory System at~DAMTP (University of Cambridge), operated on behalf of the STFC~DiRAC HPC~Facility. This equipment is funded by BIS~National E-Infrastructure Capital Grant~ST/J005673/1, and STFC Grants~ST/H008586/1 and~ST/K00333X/1. This research also used resources of the HPC~cluster~ATOCATL at IA-UNAM, Mexico and of the National Energy Research Scientific Computing Center, which is supported by the Office of Science of the U.S.~Department of Energy under Contract No.~DE-AC02-05CH11231.

We acknowledge the use of \texttt{CAMB}~\cite{Lewis:1999bs}, \texttt{CLASS}~\cite{Blas:2011rf}, \texttt{CosmoMC}/\texttt{GetDist}~\cite{Lewis:2002ah}, \texttt{IPython}~\cite{Perez:2007ipy}, \texttt{Mon-} \texttt{tePython}~\cite{Audren:2012wb}, and the Python packages \texttt{Astropy}~\cite{Robitaille:2013mpa}, \texttt{emcee}~\cite{ForemanMackey:2012ig}, \texttt{Matplotlib}~\cite{Hunter:2007mat}, \texttt{nbodykit}~\cite{Hand:2017pqn} and \texttt{NumPy}/\texttt{SciPy}~\cite{Walt:2011num}.

\clearpage
\appendix
\section{Validation of the Method}
\label{app:tests}

We have advocated the use of a phase template to characterize the effect of neutrinos. This is a natural choice as the phase shift is the physical effect we wish to isolate. It was shown in~\cite{Baumann:2017gkg} that this approach captures essentially all of the information in the BAO~spectrum at the sensitivity levels of the BOSS~experiment. However, one may still worry that the mapping\vspace{-2pt}
\beq
O_\mathrm{lin}(k) \to O^\mathrm{lin}_\mathrm{fid} \big(k/\alpha + (\beta-1) f(k)/r_d^\mathrm{fid}\big)	\label{eq:model_Neff0}\vspace{-3pt}
\eeq
introduces additional unphysical changes to the BAO~spectrum. Since we use $\Neff = 3.046$, corresponding to $\beta=1$, as the fiducial model, a poor modeling for $\beta \neq 1$ could lead to artificially strong evidence for a phase shift and could bias the determination of $\beta$ if $\Neff \neq 3.046$.
\begin{figure}[h!]
	\begin{center}
		\includegraphics{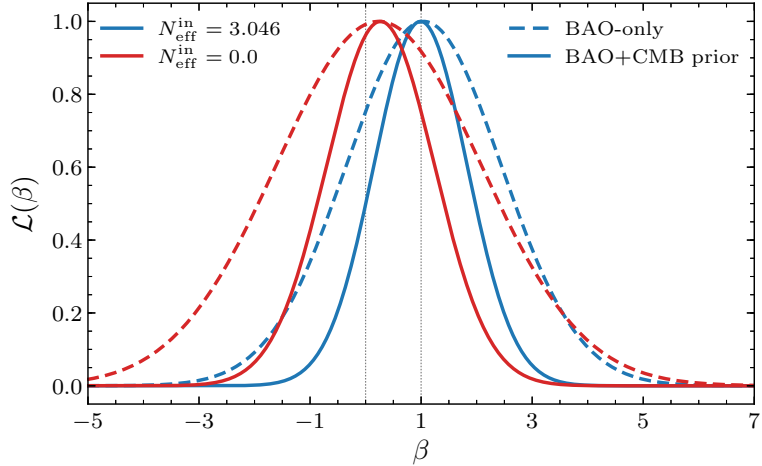}\vspace{-4pt}
		\caption{Validation of the modified BAO~analysis employed in this paper. The displayed posterior distributions for the amplitude of the phase shift~$\beta$ are computed in likelihood-based forecasts for scenarios in which the mock BOSS~data were generated using $N_\mathrm{eff}^\mathrm{in} = 3.046\text{ (\textcolor{pyBlue}{blue}) and }0$ (\textcolor{pyRed}{red}), corresponding to $\beta = 1\text{ and }0$. In both cases, the model in~\eqref{eq:model_Neff0} used a fiducial cosmology with $\Neff = 3.046$. The dashed lines show the posterior distributions after imposing a prior from a Planck-type CMB~experiment. We see that the posteriors reproduce the expected behavior which indicates that the estimation of~$\beta$ is essentially unbiased.} \vspace{-4pt}
		\label{fig:validation}
	\end{center}
\end{figure}

Our interest lies mostly in the exclusion of $\beta = 0$. A straightforward check that our method is reliable is to compute the posterior distribution for~$\beta$ in a cosmology with $\Neff = 0$ to see that the result is effectively unbiased. We use the same likelihood-based forecasts as in~\cite{Baumann:2017gkg} and the resulting posterior for~$\beta$ is shown in Fig.~\ref{fig:validation}. The expected values for~$\alpha$ and~$\beta$ are retrieved reliably in both cases. We also find good agreement when imposing the CMB~prior from Planck with the respective input values of~$\Neff$. This test demonstrates that even though the fiducial model with $\Neff = 3.046$ is used for constructing the template, the model with $\Neff = 0$ is correctly recovered.\footnote{In detail, the solid red curve in Fig.~\ref{fig:validation} shows a mean of $\bar\beta = 0.27$ rather than zero for a $\Neff=0$ cosmology. This level of bias is acceptably small given the much larger statistical error of $\sigma(\beta) = 0.97$. Of course, this bias should be accounted for when determining the precise statistical significance of the exclusion of $\beta=0$, but it does not affect our main conclusion that $\beta > 0$ at 95\%~confidence. At higher levels of sensitivity, e.g.~for~DESI, the expected values for~$\beta$ are recovered even more accurately for both $\Neff = 0\text{ and }3.046$. However, due to the smaller error bars and the slight difference between the parameter-based and template-based approaches around $\Neff=0$ for DESI~\cite{Baumann:2017gkg}, the mean $\bar{\beta}$ is found about~$0.8\hskip1pt\sigma(\beta)$ too high, whereas it is excellent for the fiducial $\Neff=3.046$.}

\vskip4pt
One may also be concerned that these results could depend sensitively on the method of BAO~extraction. Indeed, as discussed in~\cite{Baumann:2017gkg}, the phase shift template~$f(k)$ is quite sensitive to the BAO~extraction and demands a method that is accurate across a wide range in~$\Neff$. In contrast, the model in~\eqref{eq:model_Neff0} only requires an accurate BAO~extraction for the fiducial cosmology. We have verified that the results in Fig.~\ref{fig:validation} do not depend on the BAO~extraction method being used.

\section{Analysis in Configuration Space} 
\label{app:real}

\begin{figure}[h]
	\begin{center}\vspace{-4.3pt}
		\includegraphics{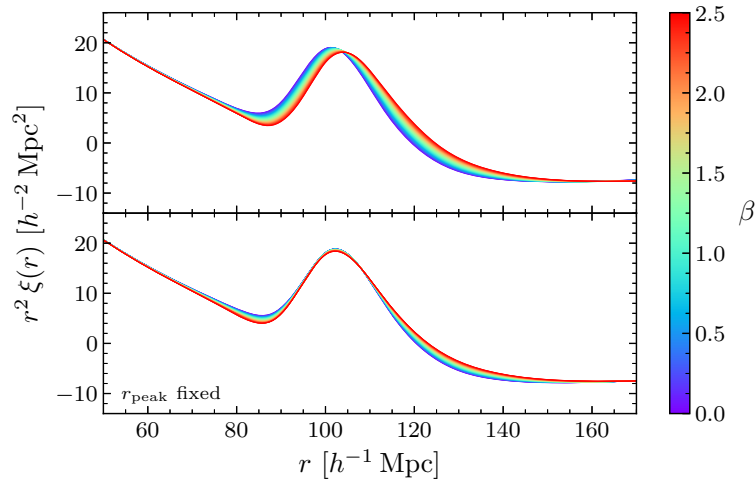}\vspace{-4pt}
		\caption{Rescaled linear correlation function~$r^2 \xi(r)$ as a function of the amplitude of the phase shift~$\beta$. The upper panel keeps the BAO~scale parameter fixed to unity, $\alpha = 1$, while~$\alpha$ is varied in the lower panel to fix the position of the peak,~$r_\mathrm{peak}$. This illustrates the degeneracy between~$\alpha$ and~$\beta$ in configuration space.}\vspace{-6pt}
		\label{fig:bao+xi_betaDependence}
	\end{center}
\end{figure}

The neutrino-induced phase shift is characteristically a Fourier-space~(FS) quantity. By contrast, the BAO~frequency is more commonly described in configuration space~(CS) as the scale of the BAO~feature in the two-point correlation function. The phase shift manifests itself in~CS as a transfer of correlations around the peak position from small to large scales (see Fig.~\ref{fig:bao+xi_betaDependence}). Given that the BAO~scale measurement is known to give compatible results in~CS and~FS (see e.g.~\cite{Alam:2016hwk}), we anticipate the same to be true of the phase shift. We will therefore implement a modified version of the CS~method used in~\cite{Vargas-Magana:2016imr} as a cross-check of our main FS~analysis.

\vskip4pt
Our nonlinear model for the correlation function starts from the processed matter power spectrum
\beq
P(k) = F(k,\Sigma_s) P_\mathrm{lin}^\mathrm{nw}(k) \left[1+O(k)\right] ,
\eeq
where~$O(k)$ is the template-based nonlinear BAO spectrum defined in~\eqref{eq:ONL} and $F(k,\Sigma_s)$~is given by~\eqref{eq:FoG}. The two-point galaxy correlation function is then modeled as
\beq
\xi_g(r) = B^2 \int \!\d\hskip-1pt\log k\, \frac{k^3}{2\pi^2}P(k)\, j_0(kr) + A(r)\, ,
\eeq
where~$j_0(kr)$ is a spherical Bessel function. We introduced the constant bias parameter~$B$ and the polynomial function~$A(r)$, taken to have the same form as in~\cite{Vargas-Magana:2016imr},\vspace{-2pt}
\beq
A(r) = \frac{a_1}{r^2} + \frac{a_2}{r} + a_3 \, ,\vspace{-3pt}
\eeq
where the coefficients~$a_n$ are marginalized over. While the constant bias matches the same parameter in the FS~analysis, the polynomial~$A(r)$ is \textit{not} equivalent to the polynomial~$A(k)$ in~\eqref{eq:fs_poly}. This is one of the notable differences between the FS~and CS~analyses. Except for the amplitude of the phase shift~$\beta$, all parameters are redshift dependent. Since the scale~$\Sigma_s$ is held fixed to the best-fit value obtained on the mock catalogs, we fit the following 13~parameters to the correlation function in the range $r\in\SIrange[open-bracket=[, close-bracket=],]{55}{160}{\per\h\Mpc}$:\footnote{We employ flat priors on the cosmological parameters, requiring~$\beta$ to be between $-10$ and~$10$, and $\alpha_z$~to be between~$0.5$ and~$1.5$, but do not impose explicit priors for the other ten parameters. On the data, we speed up the analysis by analytically marginalizing over the broadband parameters~$a_{n,z}$ in each step.}\vspace{-3pt}
\beq
\beta,\, \alpha_{z_1},\, \alpha_{z_3};\, \{B_z,\, \Sigma_{\mathrm{nl},z},\, a_{n,z} \}_{z_1,z_3} \,,\vspace{-4pt}
\eeq
for the same two redshift bins as in Fourier space.

\begin{figure}
	\begin{center}
		\includegraphics{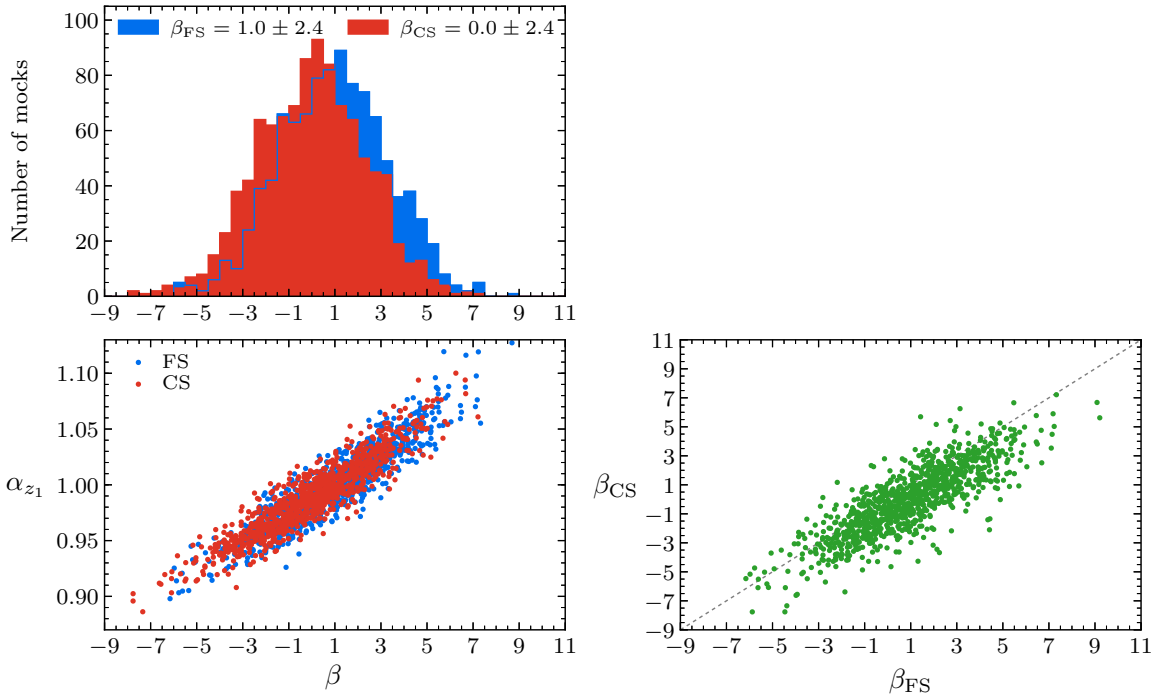}\vspace{-7pt}
		\caption{Validation of the configuration-space analysis using mock catalogs. The left column contains a comparison of the distribution of maximum-likelihood values in the \mbox{$\alpha$-$\beta$}~plane for the redshift bin~$z_1$~(\textit{bottom}) and for $\beta$~(\textit{top}) in 999~mock catalogs~\cite{Kitaura:2015uqa} for the Fourier-space~(FS, \textcolor{pyBlue}{blue}) and configuration-space~(CS, \textcolor{pyRed}{red}) analyses. On the right, we show the correlation between the inferred phase shift amplitudes in the two analyses~(\textcolor{pyGreen}{green}).}\vspace{-14pt}
		\label{fig:mocks_cs}
	\end{center}
\end{figure}
\vskip4pt
We apply the same pipeline as in~\cite{Vargas-Magana:2016imr} to the MultiDark-Patchy mock catalogs~\cite{Kitaura:2015uqa} and determine the distributions of maximum-likelihood values for the parameters~$\alpha$ and~$\beta$. The results are shown in Fig.~\ref{fig:mocks_cs} and correspond to $\beta_\mathrm{CS} = 0.0 \pm 2.4$ ($\alpha_{z_1} = 0.989 \pm 0.033$, $\alpha_{z_3} = 0.990 \pm 0.034$). Comparing these distributions with the FS~analysis of~\S\ref{sec:analysis}, we observe a strong correlation with correlation coefficient $r=0.84$, but a statistically significant bias of about $1/3$~of a standard deviation for both~$\alpha_i$ and~$\beta$, albeit with approximately the same standard deviations. When including the CMB~prior, the mean shifts upwards and gives $\beta_\mathrm{CS} = 0.75 \pm 0.89$, corresponding to a bias of about $1/4$~of a standard deviation, which is also slightly larger than in~FS. These values demonstrate good statistical agreement between the CS~and FS~analyses, and demonstrate that~CS provides a useful cross-check of the FS~analysis. While~CS does show larger biases, they are sufficiently small that they should not meaningfully affect the statistical significance of our results. On the other hand, we noticed that the precise choice of the broadband polynomial~$A(r)$ altered both the mean and standard deviation, while being consistent with the fiducial cosmology. These features of the CS~analysis will be explored in future work. The shifts seen in~CS further highlight the remarkable robustness of the phase shift in~FS.

\begin{figure}[t]
	\begin{center}
		\includegraphics{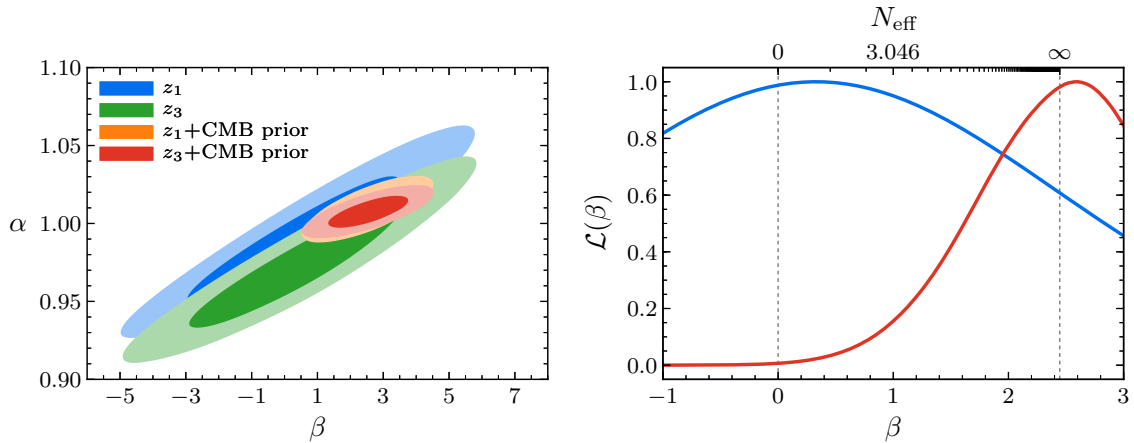}
		\caption{Observational constraints on the amplitude of the phase shift~$\beta$ from our configuration-space analysis of the BOSS~DR12~data. \textit{Left:}~Contours showing $1\sigma$~and $2\sigma$~exclusions in the plane spanned by the BAO~scale parameter~$\alpha$ and the phase shift amplitude~$\beta$ for the two redshift bins~$z_1$ and~$z_3$, both from the BAO~data alone and after imposing a CMB~prior on~$\alpha$. \textit{Right:}~One-dimensional posterior distributions of~$\beta$ without~(\textcolor{pyBlue2}{blue}) and with~(\textcolor{pyRed2}{red}) the $\alpha$-prior from the Planck satellite for the combined redshift bins resulting in $\beta_\mathrm{CS} = 0.4\pm 2.1$ and $\beta_\mathrm{CS} = 2.55 \pm 0.80$, respectively. The shift in the mean value originates from lower values of~$\alpha$ in conjunction with the discussed degeneracy between~$\alpha$ and~$\beta$.}\vspace{-7pt}
		\label{fig:2Dlike_cs}
	\end{center}
\end{figure}

\vskip4pt
With these caveats in mind, we apply the CS~pipeline to the BOSS~DR12~dataset. The posterior distributions for the parameters~$\alpha_{z_1}$, $\alpha_{z_3}$ and~$\beta$ are presented in Fig.~\ref{fig:2Dlike_cs},%
and correspond to measurements of $\alpha_{z_1} = 0.991 \pm 0.027$, $\alpha_{z_3} = 0.973 \pm 0.026$ and $\beta_\mathrm{CS} = 0.4\pm 2.1$. These mean values of~$\alpha_i$ are about $1/4$~of a standard deviation lower than the ones found in the standard BAO~analysis~\cite{Vargas-Magana:2016imr}. In addition, the error bars increased, mainly related to the degeneracy between~$\alpha$ and~$\beta$ discussed in~\S\ref{sec:analysis}. The value of~$\bar\beta$ is $0.3\hskip1pt\sigma$~lower than in~FS with a 16\%~larger error. When adding a Planck prior to break the degeneracy, we find $\beta_\mathrm{CS} = 2.55 \pm 0.80$ which is larger than in~FS because of the mentioned bias in~$\alpha_i$ towards lower values. Nevertheless, these CS~constraints are statistically consistent with the main FS~results, with similar shifts in the mean values as observed in the mock analysis. Given that the broadband modeling and peak isolation in configuration and Fourier space is distinct, an agreement between the two analyses was not guaranteed, although the change to the BAO~peak is simply the inverse Fourier transform of the phase shift. Having said that, despite these differences, this analysis confirms that a constraint, which is comparable to the main analysis in Fourier space, can also be inferred in configuration space.

\clearpage
\phantomsection
\addcontentsline{toc}{section}{References}
\bibliographystyle{utphys}
\bibliography{references}

\end{document}